\documentclass[aps,prl,twocolumn,groupedaddress]{revtex4}
\usepackage{graphicx}
\usepackage{amsfonts}
\usepackage{amssymb}
\usepackage{amsmath}
\usepackage{epstopdf}
\usepackage{verbatim}
\begin{document}

\title{Conformation of Circular DNA in 2 Dimensions}

\author{G. Witz}
\affiliation{}

\author{K. Rechendorff}
\affiliation{}

\author{J. Adamcik}
\affiliation{}

\author{G. Dietler}
\affiliation{Laboratoire de Physique de la Mati\`{e}re Vivante, Ecole Polytechnique F\'ed\'erale de Lausanne (EPFL), CH-1015 Lausanne, Switzerland}
\date{\today}

\begin{abstract}
The conformation of circular DNA molecules of various lengths adsorbed in a 2D conformation on a mica surface is studied. The results confirm the conjecture that the critical exponent $\nu$ is topologically invariant and equal to the SAW value (in the present case $\nu=3/4$), and that the topology and dimensionality of the system strongly influences the cross-over between the rigid regime and the self-avoiding regime at a scale $L\approx 8\ \ell_p$. Additionally, the bond correlation function scales with the molecular length $L$ as predicted. For molecular lengths $L\leq5\ \ell_p$, circular DNA behaves like a stiff molecule with approximately elliptic shape.
\end{abstract}
\maketitle 
From a theoretical and experimental perspective, the physics of linear polymers is well-established \cite{degennes} and its predictions were also confirmed for linear DNA \cite{maier, valle,yoshinaga,witz}. On the contrary, topologically constrained polymers, like circular or knotted DNA, are still challenging, and the topological effects on the static and dynamic properties are a matter of debate \cite{descloizeaux,duplantier,fisher_2D,grosberg,frey,orlandini, dobay}. A convergence has been reached on the theoretical value of the critical exponent $\nu$ describing the divergence of the radius of gyration with the length of circular and knotted polymers \cite{duplantier,jagodzinski,deguchi}, and fixes this value to the SAW exponent. Recently, experimental evidence has been found that the scaling exponent $\nu$ for knotted DNA in 3D corresponds to the 3D linear SAW value \cite{ercolini}. For circular DNA in 3D, diffusion measurements have pointed in the same direction \cite{robertson}. Real polymers, like DNA, have a mechanical stiffness which strongly influences the scaling properties as well as the bond correlation function, making the theoretical treatment more complex. Semiflexible polymer models were put forward in order to include this into the theory \cite{wilhelm_frey,winkler,thirumalai}, and showed the existence of a gaussian regime between the rigid and the self-avoiding regimes \cite{yoshinaga}. Many theoretical investigations concerning the properties of circular molecules were motivated by the observation of DNA in closed form \cite{zimm,bloomfield,croxton,baumgartner,jang}. These studies are relevant for biology because DNA frequently adopts circular form {\it in vivo} \cite{bates}. Furthermore, the large persistence length $\ell_p$ of DNA permits to test semi-flexible polymer theories and the effect of correlation along the molecule.

In this Letter, we present the first results for circular DNA in 2D spanning lengths from $L/\ell_p = 2$ up to $120$, where $L$ is the plasmid length.  We show that the gaussian regime is suppressed in 2D in favor of a direct transition from the rigid rod behavior to the self-avoiding walk behavior indicating a strong effect of the dimensionality and of the topology on this transition \cite{fisher_2D}. The scaling of $\langle R_G^2 \rangle$ is confirmed, and the internal end-to-end distance $\langle \xi^2 \rangle$ is well-described by the Bloomfield-Zimm formula \cite{bloomfield}. The correlation function scales as predicted by theoretical studies \cite{baumgartner,croxton} with the critical exponent for a SAW in 2D. For the short plasmids, $L< 5\ \ell_p$ a rigid rod behavior for the tested quantities was observed.
\begin{figure}
  \includegraphics[width=7.8 cm]{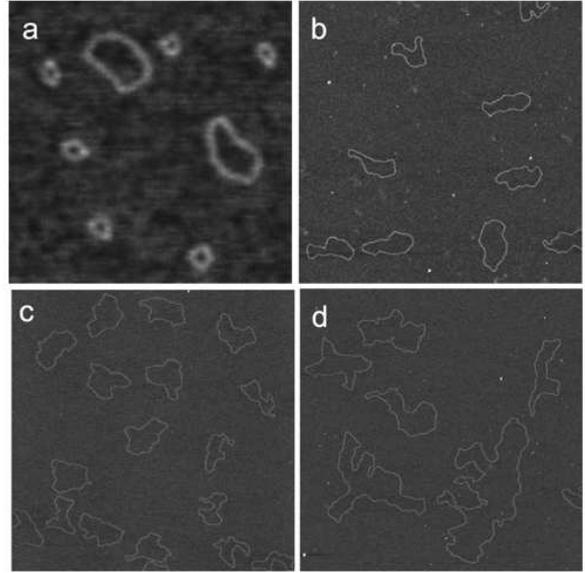}
  \caption{AFM images of DNA plasmids deposited on Mg$^{2+}$-mica. (a) minicircle-1 and -2 (350 nm scale) (b) pUC19 (2 $\mu$m scale), (c) pBR322 (3 $\mu$m scale), (d) pSH1 monomeric, dimeric, and trimeric forms (2.5 $\mu$m scale).}
  \label{afmimage}
\end{figure}

Plasmids pUC19 (2686 bp), pBR322 (4361 bp), and pSH1 (5930 bp) were
nicked to produce circular molecules without superhelicity. Plasmid pCU19 was treated with restriction enzyme RsaI to produce 3 different linear fragments, and using T4 DNA ligase mini-circles of different length were obtained. All DNA
preparations were diluted in 1 mM Tris-HCl buffer, pH 7.8, to a final DNA
concentration of 1 $\mu$g/ml. MgCl$_{2}$ was added in the DNA
solution to a final concentration of 5 mM. A 10 $\mu$l aliquot of
the DNA solution was deposited onto freshly cleaved mica, and
incubated for 10 minutes at room temperature. The sample was then
rinsed with nanopure (Ultra High Quality) water (USF Elga, High
Wycombe, England) and dried with air. Images were collected using a
Nanoscope IIIa (Veeco Inc, USA) operated in tapping
mode in air. Ultrasharp, non-contact silicon cantilevers (NT-MDT Co.,
Zelenograd, Moscow, Russia) with a nominal tip radius of $<$10 nm
were driven at oscillation frequencies in the range of
150 to 300 kHz. During imaging, the surface was scanned at a rate of
one line per second. Images were simply flattened using the
Nanoscope III software, and no further image processing was done. The
contour of the DNA molecules was extracted with \textit{Ellipse}
\cite{marek}.

From the AFM images shown in Fig. \ref{afmimage}, it is readily
clear that the molecules exhibit no self-crossings, indicating that
they equilibrate in a 2D conformation. This finding is
supported by an analysis of the persistence length, $\ell_p$, using the
bond-correlation function for semi-flexible polymers in 2D:
\begin{equation}
\langle \cos \theta (s) \rangle = \exp(-s/2\ell_p), \label{persistence}
\end{equation}
where $\theta$ is the angle between the tangent vectors to the chain at two
points separated by the distance $s$. A proper fitting can only be done for molecules with $L\geq 10\ \ell_p$ \cite{additional}, below this length, the mechanical stiffness imposes a disk shape. Fitting the experimental correlation function with Eq. (\ref{persistence}) for small separations $s$, values of
$\ell_p$ close to $50$ nm for pUC19, pBR322 and pSH1 (see Table \ref{tab_1}) were found, in good agreement with the one for linear DNA \cite{valle}, indicating that the molecules do adopt an equilibrated 2D conformation. Using the 3D equivalent to Eq. (\ref{persistence}) one
would overestimate $\ell_p$ by a factor of $2$ \cite{stasiak_2,valle}.

\begin{figure}
  \includegraphics[width=7.8 cm]{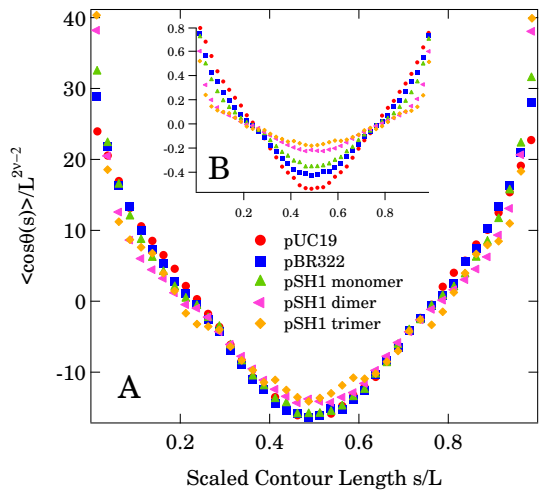}
  \caption{(a) Bond correlation function scaled according to Eq. (\ref{baum}) with $\nu$ = 0.75. (b) Unscaled bond correlation function.}
 \label{corr}
\end{figure}

In Fig.  \ref{corr}B the bond correlation functions for the entire range of segment separations $s$ are depicted for plasmid lengths $L\geq\ 10\ \ell_p$. Unlike the case of linear chains, after the initial exponential decay, circularity imposes negative values with a symmetrical minimum at $s\ =\ L/2$. In the case of a long flexible chain, the bond correlation function was studied in several theoretical papers \cite{baumgartner, croxton}. For a random walk, the correlation would be uniformly distributed over all segments \cite{croxton}. However, self-avoidance alters this distribution, and farther spaced segments will, on average,  have a more negative correlation depending on the total length. A scaling relation for the bond correlation function was suggested by Baumg\"{a}rtner \cite{baumgartner};
\begin{equation}\label{baum}
    \langle \cos \theta(s) \rangle = \phi \left( \frac{s}{L},v
    \right) L^{2\nu-2},
\end{equation}
where $\phi$ is a scaling function depending on the relative length $s/L$,
$v$ the excluded volume, and $\nu$ the critical exponent
pertaining to the scaling of the radius of gyration. Rescaling our
data according to Eq. (\ref{baum}), and using the critical exponent for a SAW in 2D ($\nu$ =
0.75), the data collapse on one universal curve as shown in Fig. \ref{corr}A.
The fact that $\nu$ = 0.75 results in a good data collapse is a further confirmation of the 2D equilibration of the DNA plasmids on long length scales.

\begin{figure}
  \includegraphics[width=7.8 cm]{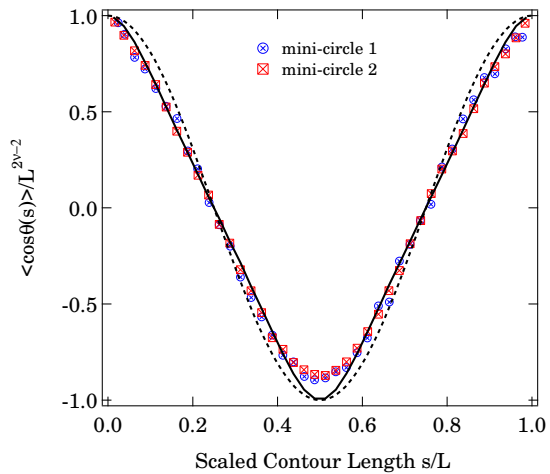}
  \caption{Bond correlations for the smallest molecules. The theoretical correlation function for a circle (dashed) and for an ellipse (continous) with eccentricity 0.67 are also drawn.}
 \label{corrmini}
\end{figure}

For the two smallest molecules ($L\ =\ 1.6\ \ell_p$ and $4.6\ \ell_p$), the correlation function (depicted in Fig. \ref{corrmini}) is closer to that of  an ellipse of eccentricity 0.67 \cite{frey}. A small deviation around $s=L/2$ still remains, whose origin is not yet clear \cite{frey,widom}.
 
\begin{figure}
  \includegraphics[width=7.8 cm]{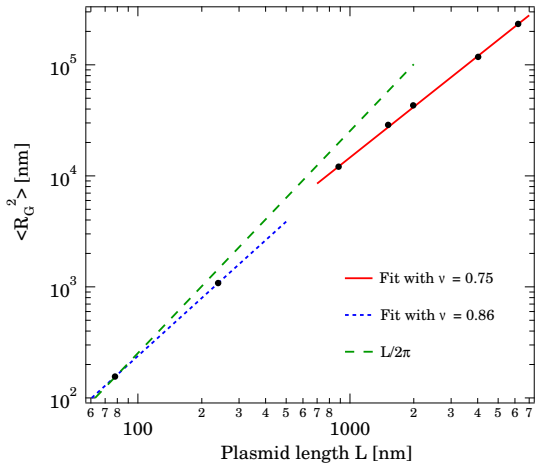}\\
  \caption{Radius of gyration as a function of the plasmid length. Fitting with $\langle R_G^2 \rangle \sim L^{2\nu}$ gives $\nu = 0.86\pm 0.03$ for the small scale regime and  $\nu = 0.75 \pm 0.04$ for the large scale regime.}\label{gyrrad}
\end{figure}

Next, we determined the radius of gyration $\langle R_G^2 \rangle$ as a function of the molecular length $L$ in order to test the scaling relation $\langle R_G^2 \rangle \sim L^{2\nu}$. To ensure good statistics, the average for each molecular length was taken over about 100 molecules (except the case of pSH1-trimer with only 25 molecules). In Fig. \ref{gyrrad} we show a log-log plot of $\langle R_G^2 \rangle$ vs. $L$ with power law fits of the data, giving directly the value of the critical exponent $\nu$. Two distinct regimes appear on this graph. A fit through the two data points for smallest $L$ gives a value of $\nu = 0.86\pm 0.03$, expressing the fact that for these almost circular molecules the radius of gyration $R_G$ is equal to the radius of a circle, thus $R_G = L/2\pi$. In Fig. \ref{gyrrad} the dashed line represents the radius of gyration $\left<R_G^2\right>$ for a perfect circle. For the larger molecules, we obtain a value $\nu = 0.75 \pm 0.04$, in very good agreement with the theoretical value for a SAW in 2D. This clearly shows that circular molecules exhibit scaling with molecular length, and that the scaling exponent is topologically invariant.

\begin{table}
\centering \caption{Properties of the investigated DNA molecules.
The length $L$ is calculated using 0.34 nm per basepair. All lengths are in nanometers.}
\begin{tabular}{cc|c|c|c|c|c}
  \hline
   DNA type & (bp) & $L$ & $L/\ell_p$ & $\ell_p$ & $L_{\textrm{fit}}$ & $\nu$ \\
   \hline
   mini $\sharp$1 & 241 & 82 & 1.6 & N.A. & 83$\pm$1 &
   0.90$\pm$0.01 \\
   mini $\sharp$2 & 676 & 230 & 4.6 & N.A. & 241$\pm$1 &
   0.91$\pm$0.01 \\
   pUC19 & 2686 & 913 & 18.3 & 52.5$\pm$5.1 & 937$\pm$13 & 0.85$\pm$0.07 \\
   pBR322 & 4361 & 1483 & 30 & 49.1$\pm$4.6 & 1543$\pm$5 & 0.82$\pm$0.01 \\
   pSH1 & 5930 & 2016 & 40 & 53.5$\pm$4.2 & 2110$\pm$10 & 0.81$\pm$0.01 \\
   2 x pSH1 & 11860 & 4032 & 80 & 49.1$\pm$2.8 & 4125$\pm$12 & 0.76$\pm$0.01 \\
   3 x pSH1 & 17790 & 6048 & 120 & 49.6$\pm$2.5 & 6293$\pm$18 & 0.75$\pm$0.01 \\
  \hline
\end{tabular}\label{tab_1}
\end{table}

Fig. \ref{gyrrad} also suggests that there is no gaussian regime ($\nu=1/2$) between the stiff and the SAW regimes as it was found for linear polymers \cite{yoshinaga}. This behavior could be due to the topology and the dimensionality of our DNA. In fact, Camacho et al. \cite{fisher_2D} described the scaling properties of semi-flexible circular polymers in 2D, and did not find an intermediate gaussian regime. The reasons are that collisions between two strands are more likely in 2D, and that circularity imposes to the polymer strand to come back on itself: both effects increase the self-avoiding interaction on short distances. From Fig. \ref{gyrrad} the cross-over length (= the length at which the self-avoidance manifests in the radius of gyration) was estimated by locating the crossing of the lines with slope $\nu=1$ and $\nu=3/4$. This gave $L=400$ nm $\simeq 8\ell_p$, agreeing well with the results in Ref. \cite{fisher_2D}.

\begin{figure}
  \includegraphics[width=7.8 cm]{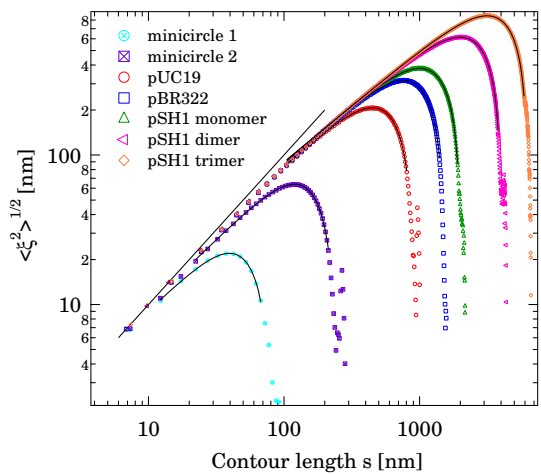}
  \caption{The internal end-to-end distance $\sqrt{\left<\xi^2\right>}$ as a function of contour length $s$ for seven lengths of DNA plasmids.
  Fits are according to Eq.
  (\ref{ete}), and the results are summarized in Table \ref{tab_1}.}\label{endtoend_1}
\end{figure}

An alternative method to determine the critical exponent $\nu$ uses the
separations of segments within individual molecules. Although this quantity was measured for linear chains \cite{valle}, hardly any experimental data exist for ring polymers. Numerical values for $\nu$ were determined by Ref. \cite{deguchi}. For a
linear molecule the scaling of intrachain separations equals that of
the end-to-end distance, and thus a larger contour length implies a
larger separation. However, for a circular molecule the topological
constraint limits the intrachain separations to $L/2$, and a specific expression
must be used. For a ring polymer with excluded volume effects, the internal end-to-end distance $\langle \xi^2(s) \rangle$ is given by
\cite{bloomfield}:
\begin{equation}
\langle \xi^2(s) \rangle \sim
\frac{s^{2\nu}(L-s)^{2\nu}}{s^{2\nu}+(L-s)^{2\nu}}.\label{ete}
\end{equation}
where $s$ the contour length. The data for
the intrachain separations are shown in Fig. \ref{endtoend_1}. Several
interesting observations can be made. As the molecules are
circular, the value of $\langle \xi^2 \rangle$ increases only up to
contour lengths $s \simeq L/2$. For larger values, the end-to-end
distance decreases back to zero. In the case of the largest molecules, for
contour length below $\ell_p$, a line with slope one describes the data well, the
reason being that the molecules on this length scale are rod-like.
Increasing the contour length beyond $\ell_p$, the self-avoiding regime is entered. Here, the molecules are highly
flexible, and are well-described by SAW walk
with circular constraint. Fitting this large-intrachain
separation regime with Eq. (\ref{ete}), we obtain values for the
critical exponent $\nu$, and the total contour length $L$ as shown
in Table \ref{tab_1}. For the longest molecules good
agreement with the expected molecular length as well as the
exponent is found. However, for short molecular lengths the fitted value of
$\nu$ deviates significantly from 0.75, while $L$ still is in good
accord with the expected length. As the relative importance of $\ell_p$ to $L$ increases for small chains, the available size of the SAW regime becomes limited and the internal end-to-end distance $\xi$ is better described by the one corresponding to a circle, namely $\xi^2(s)=2(L/2\pi)^2(1-\cos(2\pi s/L))$ or with a similar expression for an ellipse. For the two smallest minicircles, only the stiff regime is visible, because the molecules are too short to enter the SAW regime and thus the critical exponent has a value of $\nu \simeq 0.9$.

The internal end-to-end distances $\langle\xi^2(s)\rangle$ is well-fitted with Eq. \ref{ete} and therefore should follow a universal curve if $\sqrt{\langle\xi^2(s)\rangle}$ divided by $L^\nu$ and the contour length $s$ by $L$. Indeed this is the case and the rescaled plot is presented in the supplementary material \cite{additional}.

In conclusion, we have shown that circular DNA molecules adsorbed on Mg$^{2+}$-mica equilibrate in 2D conformation, and we were able to confirm several theoretical predictions concerning ring polymers. The bond correlation function obeys well to the predicted scaling \cite{baumgartner} within reasonable error when long plasmids are considered $L>>\ell_p$: this scaling seems to be reached only for the two longest plasmids ($L=80\ \&\ 120\ \ell_p$). In particular the bond correlation function (not presented here, but in Ref.\cite{additional}) decays exponentially for contour lengths $s\leq  \ell_p$, and this leads to the determination of the persistence lengths reported in Table \ref{tab_1}. Beyond the initial exponential part, nothing is known about the exact dependence of correlation function upon $s$ for $s>\ell_p$: this dependence is not trivial as it can be seen in Fig. \ref{corr}. It must be noticed that the scaled correlation functions cross each other at the reduced length $s/L=1/\pi$. Additional theoretical studies are needed to understand this behavior, especially in the case of polyelectrolytes like DNA. For short plasmids ($L\leq5\ \ell_p$), the correlation function can be better described by that of an ellipse of eccentricity $e=0.67$, whose origin could be fluctuations \cite{frey} or flexibility \cite{widom,amzallag}. This fact rises questions about the reasons of the deviations from the circular shape and should further investigated both experimentally and theoretically \cite{frey,vologodskii_2}. Furthermore, from $\langle R_G^2(s) \rangle$ vs. $L$, a cross-over from stiff to flexible SAW behavior is observed without an intermediate gaussian regime (Fig. \ref{gyrrad}). This transition takes place at $L\simeq 8\ \ell_p$, a number that seems due to the topological constraint of circularity \cite{fisher_2D}. The critical exponent $\nu$ was found to be $\nu=0.75$, and exhibits topological invariance. Similarly, from the internal end-to-end distance $\langle \xi^2(s) \rangle$ vs. $s$ (Fig. \ref{ete}) we confirm the predictions by Ref. \cite{bloomfield}. Interestingly Eq. \ref{ete} is valid for stiff as well as flexible circular molecules.  In summary, the available plasmid lengths permitted to cover a large range of conformations, from totally stiff chains to SAW chains. The importance of stiffness was shown on molecules with a length of a few $\ell_p$.

\begin{acknowledgments}
The authors acknowledge S. M. Lewis for providing the pSH1 plasmids and P. De Los Rios, B. Duplantier, R. Metzler, and A. Stasiak for discussions. G.D. thanks the Swiss National
Science Foundation for support through Grant No. 200021-116515. K.R. acknowledges the Carlsberg Foundation for financial support.
\end{acknowledgments}

\end{document}